# TERRESTRIAL ORIGIN OF SNC METEORITES AND SHOWER SOURCE FOR 30 MYR EXTINCTIONS

E.M.Drobyshevski

*A.F.Ioffe Physical-Technical Institute, Russian Academy of Sciences,
194021 St.-Petersburg, Russia (E-mail: emdrob@pop.ioffe.rssi.ru)*

**Abstract.** The inclusion of the strong strengthening of material under the conditions of high confining pressure and of the jet nature of gas outflow from impact craters has permitted us to show that meteoroid impacts of energy $\geq 10^6$ Mt TNT equivalent are capable of ejecting fragments up to ~1 km in size from the Earth into space. Accidental continental impacts of such an energy occur, on the average, once every 25-30 Myr and cause a full-scale global nuclear winter. The infall of the ejected asteroid-size fragments back onto the Earth during the subsequent 1-3 Myr would produce a series of "minimal" nuclear winters, thus accounting for the stepwise character of mass extinctions at the boundaries between some geological epochs. The possibility of ejection from the Earth of rocky fragments (including rocks of the flood basalt volcanism) permits one to propose the terrestrial rather than Martian origin of the shergottites, nakhlites and Chassigni (SNC meteorites) which removes many difficulties (an expected ratio of SNC/lunar meteorite masses opposite to the observed figure, the absence on Mars of a crater that could be associated with the ejection, the presence of terrestrial-type organics etc.).

It is shown that the isotopic anomalies of nitrogen, argon and xenon traditionally attributed to gases trapped by SNC meteorites from the Martian atmosphere can be related with terrestrial mantle samples. The same applies to the oxygen isotope shifts if one recalls the numerous known manifestations of the kinetic effects of isotopic fractionation in terrestrial rocks. Apart from this, a mechanism is proposed to explain the various oxygen isotopic shifts observed in meteorites of various types which is based on the assumption of a large difference between the ion mobilities of different oxygen isotopes in rocks due to differences in the collective behavior of $^{16}$O (which forms a continuous network in a crystalline lattice) and low-abundant $^{17}$O and $^{18}$O (present as discrete ionic inclusions). In a gas-dynamic ejection of SNC meteorites by a hot jet of gas, predominantly oxygen, the $^{16}$O ions percolate through the crystal lattice of the accelerated fragments, whereas the $^{17}$O and $^{18}$O ions become stuck and accumulate in it, thus accounting, possibly, for the observed isotopic shifts.

# I. INTRODUCTION. THE PROBLEM OF METEORITES AND OF METEOROID SHOWERS FOR MASS EXTINCTIONS

The bulk of the meteorite falls is made up by chondrites ($\approx 85\%$), with the remainder divided among achondrites (7-8%), irons (6-7%) and rock-irons (1-1.5%) (McSween, 1987). They primarily come from the Near-Earth Asteroids (NEAs) (Simonenko, 1979). 92 such objects were listed by Spratt (1987): six in the Aten group (with orbits lying inside that of the Earth), 40 in the Apollo group with perihelia inside Earth's orbit, and aphelia outside it, and 46 in the Amor group with perihelia touching it. As of September 1990, 143 of such objects have been discovered altogether, and they are revealed, on the average, at the rate of one asteroid a month, primarily due to efforts of E.Helin with coworkers (Asteroid Update, 1990). It is the NEAs that provide, on the other hand, $\approx 82\%$ of the total present-day bombardment of the Earth responsible for craters with diameters $D = 10$ km, whereas short-period comets account for $\approx 12\%$, and long-period ones, for $\approx 6\%$ (Olsson-Steel, 1987; Weissman, 1982). Nevertheless, the lack of meteorite sources is a traditional problem. If the NEAs are not replenished, their population should be in excess of $10^2 M_{Sun}$-$10^3 M_{Sun}$ initially (McSween, 1987; Opik, 1963). However the diffusion of the asteroids in the course of their mutual collisions and planetary perturbations does not provide the required number of the NEAs. The last effort in removing this difficulty was made by Wetherill (1988) who considered the evolution of fragments due to asteroid collisions in a region distant less than 2.6 AU from the Sun. Adopting a figure of 3% for the fragments acquiring a velocity perturbation of 100 m/s, an assumption bordering on the limits of credibility, he came to the conclusion that this process could remove the contradiction. Nevertheless, he pointed out that "best estimate" of the predicted figure of asteroid origin amounts only to one half of the figure derived from direct observations. Long before this, Opik (1963) suggested that part of the NEAs are actually former comets. Wetherill (1988) also inclines to this opinion. Encke's comet with its quasi-asteroidal orbit was always an inspiring argument for this opinion. A difficulty arises here (Simonenko, 1979) if one remains within the standard condensation-sublimation concepts of the origin of comets, since most of the meteorites are metamorphosed chondrites, which requires the presence in the cometary nucleus of an internal energy source to heat their rocky matrix. Besides, in this context the origin of objects of the type of Encke's comet is likewise unclear; indeed, its aphelion does not reach by far the Jupiter's orbit. If, however, one assumes that the ice of the short-period comets is saturated by the products of its electrolysis (Drobyshevski, 1988) making it capable of combustion and even of exploding, and that the C-type asteroids are actually superficially outgassed fragments of the icy envelope of the ice planet Phaethon ($M \leq 0.5 M_{moon}$) which exploded somewhere between Mars and Jupiter (Drobyshevski, 1986), then these difficulties (as many others in cometary physics) become removed. There is another problem that still remains, and it is the matching with the observed relative numbers of the low and high albedo NEAs: indeed, from some observations, too large fraction of NEAs have an albedo ($\approx 0.15$-$0.2$) which is high for an extinct comet, which implies that they most likely are S-asteroids (for a relevant discussion and references see Luu and Jewitt (1989)).

The problem of the lack of meteorite sources correlates with that of the sources of catastrophic shower bombardments of the Earth at the boundaries of some geological epochs marked by mass extinctions. It was Laplace (1861) who suggested that a collision of a comet with the Earth could have catastrophic consequences for its population as a result of a change in the planet's rotation. Kelly and Dachille (1953) pointed out the drastic consequences of supertsunamis resulting from oceanic impacts. A more modern analysis of what would happen if a large impactor struck the Earth (see, e.g. Alvarez and Asaro (1990); Grieve (1987), and refs. therein) suggests, besides the destructive action of the supertsunamis, also the effect of strong shock waves and the resultant overheating of the atmosphere, the poisoning of the air and water basins by toxic cometary substances etc. The primary impact on life would come apparently from the influx into the atmosphere of huge amounts of submicrometer-size silicate dust and soot, as a result of which the illumination level of the Earth's surface would drop by several orders of magnitude, and the near-surface air would cool in a few weeks by 10-20 K with the advent of the so-called nuclear winter which would last for months (Toon *et al.,* 1982; Turco *et al.,* 1984).

Urey (1973) and Napier and Clube (1979) suggested that the large-scale well-documented mass extinction at the Cretaceous/Tertiary boundary 65 Myr ago when 75% of all living species disappeared (including the dinosaurs) could have resulted from a climatic catastrophe caused by an impact of a large body, namely, a comet nucleus or an asteroid. Basing on the absolute and relative abundances of iridium and other siderophylic elements in a ~2 cm thick clay layer corresponding to that period and found in different areas on the Earth, Alvarez *et al.* (1980) came to a conclusion that the impactor was 10±4 km in size. It belongs in composition to the most primitive C1-type meteorites (Alvarez *et al.,* 1980; Ganapathy, 1980). Somewhat later, however, there appeared indications that these extinctions were of a stepwise character, in other words, the major events were possibly preceded by less significant ones and occurred in several stages during 1-3 Myr (see, e.g. Hut *et al.,* 1987). At the same time several large impacts could be identified (Badyukov *et al.,* 1987).

Raup and Sepkoski (1984) discovered that during the recent 250 Myr, mass extinctions occurred with a periodicity of 25-30 Myr. The same periodicity was found to exist in the ages of large craters (Alvarez and Muller, 1984) and some other events on the Earth (Bailey *et al.,* 1990). These conclusions, however, are not unambiguous. First, the data including age estimates, with respect to both biostratigraphy and other events (impact craters etc.) are rather incomplete, this shortcoming becoming progressively more pronounced as one moves back in time (Grieve *et al.,* 1985). Second, it is unclear what should be understood, for example, by mass extinctions, since species appear and die out continuously, irrespective of large catastrophes (Newell, 1982). In the light of such uncertainties one can hardly insist on the existence of a 26-30 Myr periodicity, this being rather the mean interval between such events (Walter and Finkel, 1987; Stigler and Wagner, 1987). Nevertheless, a number of hypotheses were put forward to support the conclusion of Raup and Sepkoski. The most popular is the hypothesis involving the formation of comet showers in the hypothetical Oort cloud (Hills, 1981; Hut *et al.,* 1987). A comparison of the above figures suggests, however, that postulating an enhancement of the comet flux can hardly be justified. To increase noticeably the terrestrial impact rate which is dominated by asteroids, the number of the long-period comets should be increased by at least two orders of magnitude. This is why, in particular. Bailey *et al.* (1987) came to a conclusion that known astronomical causes cannot account for the 30 Myr mass extinctions.

And while the problem is still in the stage of active discussion (Bailey *et al.,* 1990; Sepkoski, 1989; Clube, 1989) and no common final opinion has yet formed, one can draw from the above a logical conclusion that there is a sense in searching for some other, besides those already pointed out, sources of showers which would be closer to the Earth and. preferably, of asteroidal nature.

I have recently pointed out (Drobyshevski, 1989) two other possible sources of the shower bombardment of the Earth where catastrophic impacts on the Earth are preceded, as a rule, by less catastrophic events.

The first of them involves explosion of massive electrolyzed icy envelopes on bodies of the type of Jovian Galilean satellites. These are extremely rare events (once in ~800 Myr). The last catastrophe, presumably the explosion on Ganymede, can be dated back to the Cryptozoic/Phanerozoic transition (≈570 Myr BP) when Callisto's surface suffered a massive bombardment whose traces still remain visible despite the relatively low viscosity of ices. Apparently, it is due to this event which created a reservoir of meteoroids of the type of the Trojans (see below), that the mean infall of large meteoroids onto the Earth increased by about a factor two in the recent 0.5 Gyr compared with the preceding 3 Gyr (Shoemaker and Wolfe, 1982). This event resulted in a mass extinction whose traces have only recently been revealed (Zang and Walter, 1989). Gostin *et al.* (1989) point out the existence of an iridium anomaly ≈600 Myr BP corresponding to a crater of $D » 85$ km. The recent ($\approx 10^4$ years ago) explosion on Titan (Drobyshevski, 1981) while possibly produced a 100-fold increase in the number of short-period comet (Clube and Napier, 1982) still has not caused a massive infall onto the Earth because of the remoteness of Saturn system, its screening by the Jupiter and a short lifetime of bodies between the Jupiter's and Saturn's orbits (~$10^4$ yr by Lecar and Franklin (1973)).

The second source which actually derives from the first one involves the explosion of the electrolyzed ices probably still present deep inside the large (≥50 km) Trojans and in this way creates *short-period comet showers*. In the present epoch it also becomes active rather infrequently, once in ~$10^8$ yr.

I have also pointed out (Drobyshevski, 1989) that the 25-30 Myr interval could correspond to purely stochastic continental impacts of energy $W \geq 10^6$ Mt TNT, which exceeds by about an order of magnitude the energy acquired for the onset of the "minimal" nuclear winter ($W \sim 10^5$ Mt TNT). It should be said that, as this was already pointed out earlier (see, e.g., Napier and Clube, 1979; Wetheri11 and Shoemaker, 1982), the impacts, say, of ~10 km bodies, one of which apparently caused the Cretaceous/Tertiary transition, occur with an interval of ~40-50 Myr close to the mean biostratigraphic period.

We related the possibility of a mass extinction, however, with a certain reference which can be a "minimal" nuclear winter, while pointed out that only an event more powerful by an order of magnitude than this could result in a global catastrophe. The stepwise and extended character of the extinctions, the multiplicity of large impacts were attributed to the ejection of asteroid-size bodies from the Earth into space in such collisions (Drobyshevski, 1990). In impacts with energy $W \leq 10^5$ Mt TNT equivalent no such large bodies can be ejected. Therefore $W \sim 10^6$ Mt TNT events are of a threshold nature both in their direct impact on the biota and in the more remote consequences of the appearance of numerous NEAs.

Below a particular mechanism is presented when one such very strong impact results in other quite powerful impacts during the next 1-3 Myr and the major consequences of such an event are discussed.

## II. STATUS OF SNC METEORITES

Consider the problem of the so-called SNC meteorites (shergottites, nakhlites, and Chassigny, altogether 8 of ~80 kg total mass out of 2,500 non-Antarctic stone meteorites) (see reviews of McSween (1985), Laul (1986), as well as the monograph Sidorov and Zolotov (1989)). They differ so strongly in composition (basalts with a high content of volatiles) from the other achondrites that one can safely conclude that their matter survived several stages of igneous differentiation under a gravitation stronger than that of the Moon, their crystallization age being fairly short (≤1.3 Gyr). The shergottites reveal indications of shock metamorphism 180-200 Myr BP (at pressures up to 300-450 kbar), while the nakhlites do not have this feature. The trapped gases correspond to the atmospheric composition of Mars (or the Earth). The escape velocity from Mars (5 km/s) being substantially smaller than that of the Earth (11.2 km/s), they are assumed to be of Martian origin (McSween, 1987, 1985; Wasson and Wetherill, 1979). The possibility of meteorite ejection from planets is supported by the discovery of eight lunar meteorites (of total mass ≈2 kg) in the Antarctic ices.

Two major mechanisms of impact-related meteorite ejection from Mars have been proposed: (1) acceleration of fragments by a shock wave in the solid rock (Melosh, 1984; Vickery and Melosh, 1987), and (2) acceleration of fragments by the gas produced in a strong impact (which can be oblique) (Wasson and Wethrill, 1979; Nyquist, 1987; O'Keefe and Ahrens, 1986; Vickery, 1986a, 1987).

One should bear in mind that the initial size of the body determined by the time of exposure of its fragments to cosmic rays (0.5-2.5 Myr) should be ≥15 m (Wetherill, 1984). Another limitation consists in that the mass of impact-created lunar meteorites should be ~2,500 times that of the Martian ones (Wetherill, 1984), and the mass of the latter, reach ~$10^{14}$ kg (Vickery and Melosh, 1987). In actual fact, of the SNC meteorites exceeds by a factor ~40 that of the lunar ones. Taking into account the crystallization age (≤1.3 Gyr), it becomes very difficult to satisfy all these requirements for the case of Mars, since they call for ejection from a comparatively young (≈200 Myr old) crater of diameter $D = 150$-$200$ km. Such craters do not appear to exist on Mars, and there is no certainty in the presence of sufficiently young craters with $D = 100$-$150$ km (see the relevant discussion in Vickery and Melosh (1987) and refs. therein).

**Table 1**. Final velocity $V_{r\,max}$ in nondestructive ejection of boulders of diameter $\varnothing$ with $\rho_r = 3$ g/cm$^3$ having a strength $\sigma_{max} = 1$ kbar and lying on the surface at a distance $X$ from the center of impact crater in andesite with final diameter $D = 30$ km (Vickery, 1987).

| Impactor material, diameter, velocity. initial gas pressure and density in crater | $\varnothing$ (m) | $X$ (km) | $V_{r\,max}$ (km/s) | $S_{eff}$ (km) |
|---|---|---|---|---|
| Andesite 2.21 km; 20 km/s 4.7 Mbar; 6.35 g/cm$^3$ | 1 | 10 | 12.25 | 2.25 |
| | 5 | 18 | 5.06 | 1.92 |
| | 10 | 20 | 3.52 | 1.84 |
| | 15 | 22 | 2.4 | 1.30 |
| Water ice 2.71 km; 25 km/s 4.5 Mbar, 3.85 g/cm$^3$ | 1 | 17 | 14.0 | 2.94 |
| | 5 | 21.5 | 6.25 | 2.93 |
| | 10 | 25 | 3.3 | 1.63 |
| | 15 | 27 | 2.0 | 0.90 |

These considerations initiated an investigation by Vickery (1986a, 1987) of the possibility of fragment acceleration by gas ejected from craters with $D = 30$ km, although recent (0.5-2.5 Myr BP) ejections from many small Martian craters are far from being capable of producing the required SNC meteorite flux. An essential point in the investigation of Vickery is analysis of the acceleration of surface boulders by a gas flow produced in a hypervelocity impact as a result of evaporation of the target and impactor material. The impact-generated shock wave creating high pressure gradients which breaks up the material and produces fractures in (plane) surface layers of the rock does not propagate through such boulders.

Boulders of spherical shape are entrained by the gas with an acceleration (Vickery, 1986a, 1987):

$$a = \frac{3}{2} \frac{C_x}{\rho_r \varnothing} \rho_g (V_g - V_r)^2 \qquad (1)$$

where $C_x \approx 0.9$ is the gas drag coefficient at Mach numbers ≥3 (Mishin, 1967) (Vickery (1986a, 1987) assumed $C_x \approx 0.5$), $\varnothing$ is the boulder diameter, $\rho$ and $V$ are the density and velocity, respectively, the indices r and g referring to the boulder and the gas, accordingly.

The maximum acceleration that can be sustained by a boulder without destruction for a material strength $\sigma_{max}$ is

$$a_{max} = \sigma_{max} / \rho_r \cdot \varnothing \qquad (2)$$

Table 1 lists some of the results taken from the graphs of (Vickery, 1987). Here

$$S_{eff} = \rho_r V_{r\,max}^2 \varnothing / 2\sigma_{max} \qquad (3)$$

is the length of the effective acceleration path for $a = a_{max}$. For $\sigma_{max} = 1$ kbar and assuming the gas expansion after the impact to be hemispherical, Vickery (1987) obtained for the size of a body capable of overcoming the gravitational pull of Mars a figure of 5 m which exceeds considerably the results cited by others ($\varnothing = 0.1$-1 m). Note that the initial distance of the boulder from the impact axis is $X \approx 20$ km; for smaller $X$ the boulder will break up because of a too high acceleration. For $X \gg 10$-15 km, only 1 m boulders will survive, but they will gain velocities up to 12-15 km/s(!).

### III. POSSIBILITY OF HYPERVELOCITY-IMPACT EJECTION OF ASTEROID-SIZE BODIES FROM THE EARTH

The latter fact appears remarkable and deserves a closer scrutiny. Indeed, if $\sigma_{max}$ could be increased by 1.5-2 orders of magnitude, and the effective acceleration path $S_{eff}$ by one order, a possibility would open to accelerate bodies with $\varnothing \sim 10^3$ m up to velocities $\geq 11.2$ km/s, i.e. to eject asteroids from the Earth! At first glance, such an assumption may look too audacious. To make it appear more sound, one can be reminded of a serious discussion of such still larger-scale consequences of the impact as the terrestrial origin of the Moon (Ringwood, 1986) or the nearly total stripping of Mercury's silicate mantle (Benz et al., 1988).

Now, to what extent what we have said on the possibility of ejection of asteroid-size bodies from the Earth is realistic?

First, the high-temperature gas formed in the crater in a hypervelocity impact produces, according to the calculations of O'Keefe and Ahrens (1986, 1977), Basilevskyi et al. (1983), and Roddy et al. (1987), a jet normal to the target surface, rather than expanding hemispherically as this was assumed by Vickery (1986a, 1987). According to our experiments carried out on an electromagnetic railgun with solid bodies of 1 g at velocities ~5-7 km/s (Drobyshevski et al., 1990), the total jet expansion angle is $\approx \pi/4$ rather than $\pi$. Therefore the gas density in a real ejection should be an order of magnitude higher than that in a hemispherical expansion, thus resulting in a corresponding increase of the effective acceleration path length. According to the above-mentioned calculations, the transverse size and height of the jet with parameters remaining nearly constant along it are determined by the scale of $D$, i.e. the final crater size. Accepting for a rough estimation $S_{eff} \approx D/3$, then $S_{eff}$ will exceed just by a factor 3-10 that for the case of hemispherical expansion from a $D = 30$ km crater (see Table 1) (Vickery, 1986a, 1987).

Second, many basic rocks have $\sigma_{max} > 1$ kbar (diabase, 8.3 kbar; gabbro, 4.2 kbar; dunite, 2.5 kbar (Beresnev and Trushin, 1976)). Note that the strength of a material increases under a high confining pressure (Beresnev and Trushin, 1976; Meada and Jeanloz, 1990) which is used in present-day technology (e.g., Drobyshevski et al., 1991). At $p = 10$ kbar $\sigma_{max}$ is 26 kbar for diabase, 22 kbar for gabbro, and 13 kbar for dunite (Beresnev and Trushin, 1976). Quite frequently, $\sigma_{max}$ grows linearly up to $p \approx 50$ kbar (Kinsland and Bassett, 1977). In our case, where the initial pressure of the gas flowing around a body accelerated by it is on the scale of hundreds of kbar, and the time during which a load is applied to the body and it is accelerated exceeds that required for the internal stress redistribution, the desired conditions of quasistatic compression up to $p \geq 50$ kbar appear to be quite realistic. Therefore we may accept as a working hypothesis that for basalts $\sigma_{max} \approx 100$ kbar can be reached.

Assuming $S_{eff} \approx \sim D/3$, then for an impact of $W \approx 3 \cdot 10^6$ Mt TNT ($D \approx 60$ km; for dependence $D(W)$ see, e.g., Weissman (1982)) the size of a body with $\rho_r = 3$ g/cm$^3$ for $\sigma_{max} = 100$ kbar, ejected with a velocity 11.5 km/s (i.e. from the Earth), may reach, by Eq. (3), $\varnothing \approx 1$ km.

Atmosphere does not affect the ejection of bodies with $\varnothing \gg 10$ m. One can therefore expect that bodies of $\varnothing \sim 100$ m in size should likewise escape from the Earth in its collisions with meteoroid. The above assumptions can be used to make a very rough evaluation of the minimal impact power which is still large enough to cause ejection of bodies from the Earth into space.

For $\varnothing = 10^2$ m, $s_{max} = 10^2$ kbar, $r_r = 3$ g/cm$^3$ and $V_{r\,max} = 11.5$ km/s, Eq.(3) yields $S_{eff} \approx 2$ km. The size of the crater thus formed should exceed $D \approx 3 S_{eff} \geq 6$ km, which corresponds to an impact energy $W \geq 10^3$ Mt TNT. Such impacts on continents would occur with an interval ~80,000 yr (since the impactor size here is only ~0.3 km, it was assumed, in contrast to Drobyshevski (1989, 1990), that continents occupy 30% of the Earth's surface). These impacts should cause ejection only of a small number of large fragments. Interaction with the atmosphere of the jet ejected from a relatively small crater and of the accelerated fragments carried by it is a fairly complex process. Nevertheless, one may expect the tail of the smaller fragments to be effectively cut off by the atmosphere. If the jet is more powerful, and $S_{eff} \geq 8$ km, which is the height of the homogeneous atmosphere, then the gas blast from the crater will pierce effectively the atmosphere and transport into space, together with the largest fragments, many smaller ones, and it is the latter that in the subsequent 1-3 Myr will be the main suppliers of SNC meteorites. The composition of these meteorites will naturally change after each such ejection in accordance with the rocks at the impact site.

Note the boulders lying freely on the surface are not the only proper candidates for ejection from the Earth. A mountain close to the impact site can also be sheared off its foot by the shock wave propagating along the surface to form, as it were, a free boulder. If we take into consideration the inhomogeneity and layered structure of subsurface rocks, which should affect the propagation of shock waves and result in three-dimensional effects of their diffraction, interference etc., shearing off of rocks from comparatively plain regions without breakup may also become possible.

All the above provides an explanation for the appearance of unexpectedly large fragments producing secondary craters up to $D = 10$-30 km (and probably more) in size on the Moon, Mars (Vickery, 1986b), Mercury (Oberbeck *et al.*, 1977), and Ganymede and Callisto (Passey and Shoemaker, 1982). To have the corresponding energy ($10^4$-$10^5$ Mt TNT) with the low escape velocities from these bodies ($V_{esc} = 2.5$-5 km/s), the size of the monolithic fragments ejected from the primary craters should reach $\varnothing = 3$-5 km. which usually was considered impossible and caused distrust (Vickery, 1986b; Oberbeck *et al.*, 1977).

## IV. MAJOR CONSEQUENCES OF A POSSIBLE ASTEROIDS EJECTION FROM EARTH. REMOVAL OF SOME PREVIOUS DIFFICULTIES

Thus in impacts creating on the Earth craters with, say, $D \geq 50$ km (impact energy $W \geq 10^6$ Mt TNT) one may expect ejection into space of asteroid-size bodies. Note that there should exist an obvious correlation between the maximum size of the asteroid bodies and the strength of the rocks making them up. The surface of these bodies should bear the traces of ablation caused by the hot gas flowing past them.

Obviously, a final conclusion on the validity of this mechanism capable of accounting for the appearance of some asteroids in the Earth-crossing orbits can be drawn only after detailed (gasdynamic and strength) calculations of the fragment behavior in the gas jets emitted in hypervelocity impacts and, of course, after *in situ* studies.

One clearly sees, however, that the assumption of the terrestrial origin of some NEAs removes a number of difficulties:

(1) The disagreement between the observed and expected numbers of lunar and SNC meteorites (when the latter were assumed to be of Martian origin) is removed now (the attempt of Nyquist (1984) to explain the ~$10^5$-fold discrepancy between the masses of lunar and Martian meteorites as due to a fast sweeping out of the NEAs and inadequate statistics can hardly be considered convincing). The absence on Mars of a crater capable of ejecting the needed amount of SNC meteorites (Vickery, 1987) becomes understandable.

(2) The origin of the meteoroid showers needed to account for stepwise extinctions with a 25-30 Myr mean interval finds its explanation. Such showers are produced when a large number of NEAs are created in random impacts of energy $\geq 10^6$ Mt TNT. The lifetime of the NEA objects is determined by two processes (Opik, 1963; Shoemaker *et al.,* 1979): (1) infall on terrestrial planets, and (2) gravitation-assisted transfer to the Jupiter orbit from where they are eventually removed out of the Solar System. The mean Earth-collision lifetime is $\approx 25$ Myr for the still surviving Aten group bodies, $\approx 10^8$ yr for the Apollo group, and $\approx 3 \cdot 10^{10}$ yr for the Amor group (Steel and Baggaley, 1985). By the present time the orbits of these few remaining bodies underwent a profound evolution and are, as a rule, in resonance with the nearest planets and the Jupiter, i.e. are stable enough. Obviously, the lifetimes for infall to the Earth of bodies recently ejected from it into close-approach orbits will be much shorter (i.e. 1-10 Myr) which is in agreement with the extinction time scale. While the probability of the second process is 3-10 times less than that of the first (Opik, 1963), it creates bodies falling on the Earth with an energy exceeding by a few times the impact energy of recently ejected asteroids of the same mass; although this does not usually entail the formation of new asteroids it is nevertheless essential for the onset of subsequent "minimal" nuclear winters and new "extinction steps".

(3) It becomes clear why the transition layers between some geological epochs (e.g. C/T $\approx 92$ Myr BP (Orth *et al.,* 1988)) contain, besides the iridium anomaly, also elements which are more typical of the Earth's upper mantle material than of meteorites. Moreover, one may expect the appearance of NEAs as a result of a hypervelocity impact with the Moon, and a series of catastrophes initiated by their collisions with the Earth. For instance, crater Tycho ($D \approx 85$ km) was created only 138(+24,-20) Myr, and Copernicus ($D \approx 93$ km), 834(+70,-60) Myr ago (Baldwin, 1985). In a such case, an iridium anomaly may not appear on the Earth at all. Because of the different total cross sectional areas of the Earth and Moon, the probability for such an event to happen should be 20 times smaller than that involving an impact with the Earth. On the other hand, the appearance proper of a clearly pronounced global iridium anomaly caused by the infall on the Earth of a single, high-velocity and large body with a composition of primitive meteorites or a cometary nucleus appears hardly likely bearing in mind the complete vaporization of the impactor and of a part of target material and their ejection in gaseous form from the crater at a very high velocity together with numerous terrestrial rock fragments, with some of them becoming NEAs. Such is the Triassic/Jurassic no-iridium transition identified with the Manicougan crater in Quebec ($D \approx 60$ km) and having the shocked quartz in its transition layers. Iridium anomalies form apparently in shower infalls of primitive (non-differentiated) bodies appearing, for instance, in the explosions of electrolyzed Trojan-type icy bodies, when there is a sufficiently broad distribution of their fragments in mass, and large impacts are preceded and accompanied by numerous smaller collisions which affect the climate and produce chemical anomalies in sediments. In this context one can understand the appearance of local rather than global iridium anomalies at the boundaries between some epochs (Upper Devonian, Permian/Triassic etc. (Grieve, 1987)), fluctuations of the anomalies in time, as well as the climatic change long before the main event (0.2-0.3 Myr before the K/T transition (Cortillot, 1990)).

(4) The earthy origin is supported also by a comparison of mineral composition of the original tholeiitic melt of the SNC meteorites with that of the mantle and crust of the Earth and of seven different Martian models in the pressure range $0 < p \leq 30$ kbar. Coincidence is found to exist only with the Earth and the only Martian model which again is closest in composition to the Earth mantle (McSween, 1985). The ratios of the volatiles K, Rb, Cs, Tl to the refractory U are also close to the terrestrial figures.

(5) Finally, the Earth's nature explains the origin of the organic material in the EETA 79001 meteorite which was found recently by Wright *et al.* (1989) in an amount unexpectedly high to be identified with Mars while having a $^{13}C/^{12}C$ ratio characteristic of terrestrial organics.

Thus one can assume the existence of epochs when the number of NEAs increased dramatically, a large part of them being fragments ejected from the Earth as a result of continental hypervelocity impacts. Judging from the number of SNC meteoritic falls, the fraction of fragments of terrestrial origin is presently small, and they could be produced in smaller-scale impacts which did not result in global catastrophes. Nevertheless one may expect to find among the infalling meteorites pieces of terrestrial acid and sedimentary rocks, with possible remains of fossil (not Martian!) organisms; note that one and the same fragment may exhibit two melted crusts of disparate ages, one of them corresponding to the ejection from, and the other to the infall onto the Earth. A search for such samples among the Antarctic meteorites could certainly bring interesting results.

## V. ON THE NATURE OF THE ISOTOPIC ANOMALIES IN SNC METEORITES. AN ATTEMPT AT COMBATING THE HYPNOTIC EFFECT OF THE MARTIAN ORIGIN HYPOTHESIS

There are, however, a number of data which, at first glance, are difficult to correlate with the concept of the terrestrial origin of SNC meteorites.

*5.1. General Anomalies in Chemical Composition*

Indeed, if the two- or three-fold differences in the content of such relatively low-abundance elements as P, Cr, Mn and, even more so, of Cu, Ni, Cl, Br between the SNC and Earth's mantle (Wanke and Dreibus, 1988) can be attributed to incompleteness of our knowledge about the inhomogeneous composition of the mantle, or to some accidental factors, this can hardly be done for the iron abundance which in all the SNC studied exceeds by a factor two that a conventional terrestrial basalts representing products of the upper mantle depleted in iron in the course of differentiation of the Earth's core. And until meteorites with a composition identical to that of the Earth's crust rocks are found, one will have to admit that practically all SNC bodies known presently were ejected in one impact event from a region of ultrabasic volcanism geochemically uncommon for the Earth, whose rocks preserved a close-to-primitive composition not depleted in iron (e.g., Treiman, 1986). Then the differences between different SNC bodies can be accounted for by the moderate heterogeneity of such rocks on a length scale of only $\sim 10\text{-}10^3$ m. The same explanation applies also to the easiness with which one can develop a model of the hypothetical planet, i.e. the parent body of the SNC meteorites, compared to the much more inhomogeneous and complex model of the Earth and of its mantle (Treiman, 1986).

*5.2. Isotopic Anomalies of Noble Gases and Nitrogen*

Consider now the problem of isotopic anomalies in SNC meteorites. McSween (1985) pointed out that "the only evidence that ties SNC meteorites directly to Mars is the composition of gases in impact melt glasses of the EETA 79001 and ALHA 77005 meteorites". He had in mind primarily the results of the analyses which revealed an impressive coincidence of some isotopic anomalies of the gases contained in SNC glasses (Bogard and Johnson, 1983; Becker and Pepin, 1984) with those found by the Viking in the Martian atmosphere in 1974 (Nier and McElroy, 1977; Owen *et al.,* 1977).

*5.2.1. Low Efficiency of Atmospheric Gas Trapping*

As pointed out by Levskii and Drubetskoi (1988), these conclusions, however, are far from unambiguous for a number of reasons. In particular, the glass "capacity" turns out to be suspiciously large, namely, 1 cm$^3$ of glass had to trap 0.2-0.3 cm$^3$ of Martian atmosphere. Clearly enough, the trapping of gases from the much denser Earth's atmosphere could provide a more plausible explanation of their high content, all the more so that Bogard *et al.* (1984) note that the Ar/Kr/Xe ratios obtained are characteristic of both the Mars and terrestrial atmospheres.

However even here it is not all so simple as it might seem at first glance.

Indeed, in the course of being ejected by the gas jet streaming out of a hypervelocity impact crater on Mars or the Earth, when the glass forms in the melting rocks of the hurled out fragment, the fragment has simply no contact with the planetary atmosphere. It is surrounded by a very dense ($r_g \sim 1$ g/cm$^3$) and hot ($T = 10^3$-$10^4$ K) gas with the composition of the surrounding rocks! It is therefore difficult to understand the grounds for the persistence with which almost all the authors attempt to compare the composition of the gases embedded in the SNC meteorite glasses with that of the Martian atmosphere while attributing all close-to-terrestrial ratios to contamination suffered on the Earth. In this context, the experimental study by Wiens and Pepin (1988) of the shock-wave implacement of gases with initial pressure of 0.045-3 atm into basalt at downstream pressures of 200-600 kbar appears significant. The experiments yielded a discouraging result for adherents of the Martian origin viewpoint, namely, monolithic basalt absorbs only 2-6.7%, and powdered samples, 40-50%, of their volume from atmospheric environment, which is clearly less than the ~100% efficiency characteristic of one of the glass inclusions in EETA 79001. Moreover, it turns out (Wiens, 1988a) that the noble gases shock-implaced into the basalt are sited in the glass primarily (~45%) in microvesicles ($\varnothing \sim 10 \div 100$ μm), whereas in EETA 79001 only 15% of the gases is contained in them, their larger part residing most probably in inter-molecular space (Wiens, 1988b). It is also essential that the bubble gas which could probably have been trapped from the atmosphere is close in isotopic composition to the Earth's than to Mars' air.

A fully adequate modeling is naturally difficult to carry out. In particular, the basalt used differs strongly in composition from those of the SNC meteorites; the sample compression time after the passage of the shock wave in these experiments was a few microseconds whereas in ejection it can be as long as a few seconds, and shock loading can occur more than once due to generation of secondary shock waves.

*5.2.2. Noble Gases in SNC Meteorites and Earth's Mantle Rocks*

Let us see whether magmatic gases, contained, *ab initio* in rocks, rather than the atmospheric air could account for many of the observed features which are presently attributed to their capture from the Martian atmosphere.

We are going to dwell here on a comparison of the amounts of noble gases contained in SNC meteorites and, for instance, in MORB (Mean Ocean Ridge Basalt). Suffice it to say that the range of their contents in MORB (e.g. Staudacher, 1989) includes their content in SNC meteorites. It should, however, be pointed out, that the isotopic composition of argon and of other noble gases in the Earth's mantle rocks also varies within broad limits, so that it may be quite close to that of the SNC meteorites.

For instance, on the Earth the $^{40}$Ar/$^{36}$Ar ratio (which is 296 in the atmosphere) varies in rocks from ~300 to $10^5$ (and even $10^6$) (Ozima and Podosek, 1983). Moreover, in pores of even the same MORB sample this ratio can vary from 560 (large pores) to 28,000 (small pores) (Staudacher *et al.*, 1989). For the EETA 79001 glasses, the $^{40}$Ar/$^{36}$Ar ratio also varies depending on the degree of crushing at the same temperature (500 °C) from 680 to 2,000 (Wiens *et al.*, 1986). On the average, in these glasses $^{40}$Ar/$^{36}$Ar ≈ 350÷1,500 by Bogard *et al.* (1984), ≈2,000 by Becker and Pepin (1984), and Wiens *et al.* (1986), and ≈1,200 by Swindle *et al.* (1986) which might seem to correlate with $^{40}$Ar/$^{36}$Ar ≈ 3,000±1,000 for the Martian atmosphere (Owen *et al.*, 1977). However in the unmelted rocks of Shergotty and EETA 79001, $^{40}$Ar/$^{36}$Ar ≈ 700-2,000 also (Swindle *et al.*, 1986; Ott and Begemann, 1985). So that there are hardly any grounds for maintaining that the argon was trapped by the melt from the Martian atmosphere.

The terrestrial ratio $^{129}Xe/^{132}Xe$ lies from 0.983 in the atmosphere to 1.15 in some of the studied mantle rocks (Staudacher *et al.*, 1989). The main supplier of $^{129}Xe$ is assumed to have been radioactive $^{129}I$ with a decay half-period of only 16 Myr, so that the excess $^{129}Xe$ can be associated with the presence in the rocks of voids that originally contained $^{129}I$ (Ozima and Podosek, 1983).

The only Xe isotope ratio measured by the Vikings in the Martian atmosphere is $^{129}Xe/^{132}Xe$. It turned out to be 2.5(+2,-1) within large error limits (Owen *et al.*, 1977). To author's knowledge, no discussion of the reliability of these measurements has ever been carried out.

The ratios for all Xe isotopes in the SNC meteorites studied do not practically differ from those for the Earth's atmosphere (Bogard *et al.*, 1984; Swindle *et al.*, 1986; and refs. therein), except for $^{129}Xe/^{132}Xe$ which in ALHA 77005, Shergotty and EETA 79001 exceeds by 10-50% that of the air. An exclusion is the glasses in EETA 79001 where the $^{129}Xe/^{132}Xe$ ratio is ~2.4 times that for the air, which is considered, as a rule, to be an argument for the gas there being trapped from the Mars' atmosphere. So, there are some grounds, indeed, to believe that one of SNC meteorites, namely EETA 79001, is of Martian origin. We have, however, already pointed out the low efficiency of gas trapping even under favorable conditions when the samples were totally immersed in gas. It is essential also that the enriched $^{129}Xe$ is sited primarily in intermolecular space rather than in bubbles, i.e. it was present in the rocks in its original state. More probably appears the transformation $^{129}I \rightarrow {}^{139}Xe$ and preservation of the system in the closed state as is the case with some mantle rocks (Staudacher *et al.*, 1989; Ozima and Podosek, 1983; Ott and Begemann, 1985), all the more so that in EETA 79001 and ALHA 77005 the I/Br ratio exceeds by far the cosmic figure. True, Swindle *et al.* (1986), Wanke and Dreibus (1988), and Koeberl and Cassidy (1991) (and refs. therein) point out the possibility of terrestrial contamination of these Antarctic meteorites by iodine. Finally, it cannot be excluded that the gases enter the high porosity rocks transformed into glasses from neighboring reservoirs (see below). In this context, the idea of Shukolyukov and Meshik (1990) about the possibility of transport of volatile radioactive uranium decay products from the regions of Oklo-type natural fission reactors on the Earth deserves development. Their buildup in other areas (atmosphere, reservoirs in rock etc.) produces there unexpected isotopic anomalies. The authors believe the probability of the formation of such reactors to have been fairly high 2.6-2 Gyr BP, the fission-produced isotope $^{129}I$ being one of the most probably candidates for such transport and creation of relatively young reservoirs of excess $^{129}Xe$. Nevertheless, the maximum $^{129}Xe/^{132}Xe$ ratio found thus far in SNC meteorites is twice the value measured in any terrestrial rock. A discovery on the Earth of samples with $^{129}Xe/^{132}Xe \approx 2.4$ would automatically eliminate this difficulty.

Hopes for such a resolution of this problem always persist, and not without grounds, since it is likely that terrestrial rocks contain $^{129}Xe$ not in negligible amounts (Ozima, Igarashi, 1990). Indeed, Ott and Begemann (1985) point out the extreme inhomogeneity and nonequilibrium character of the noble gas distribution in terrestrial rocks (see also Ozima and Podosek, 1983) and note that atmosphere gases always contain such radiogenic nuclides as $^4He$, $^{40}Ar$, $^{129}Xe$ in smaller amounts than the rocks do. They conclude that if the gas in EETA 79001 glass was trapped from the Martian atmosphere, and "if Shergotty, Nakhla and Chassigny are also to come from Mars, then the situation on Mars must be opposite of that on Earth in that the atmosphere would have to contain a higher proportion of radiogenic xenon isotopes than does solid Mars". The same applies also to the xenon isotopes representing decay products of heavy nuclei. Such an assumption appears strange from their viewpoint for a number of other reasons as well. In particular, Ott (1988) points out some differences in the isotopic ratios in Shergotty, Nakhla and Chassigny which argue for somewhat different (i.e. nonatmospheric) original reservoirs of noble gases.

Similar conclusions are reached by Swindle *et al.* (1986) who point out difficulties associated with the trapping of noble gases from the Martian atmosphere. Just as other authors, they stress the remarkable closeness of the Kr/Xe ratio to the figure typical of the Earth's atmosphere (see also Becker and Pepin, 1984), as well as the difference in the direction of fractionation of Xe, and all the more so, of Kr, on the Earth and in SNC meteorites from that of solar or chondrite types.

*5.2.3. Enrichment by Heavy Nitrogen*

Swindle *et al.* (1986) drew attention to a remarkable noble gas component present in EETA 79001 glass and not found anywhere before in the Solar System. Possibly, it is only a pseudocomponent representing different gases mobilized from the surrounding rocks by a shock and embedded in the shock-produced glass. They do not even exclude the possibility that this redistribution was accompanied by mixing of various components and *intense isotopic fractionation*. However these authors point out at the same time that the nitrogen isotope anomalies revealed in EETA 79001 glasses (they are high, reaching for some temperature fractions the following levels with respect to air: $\delta_{air}^{15}N = [(^{15}N/^{14}N)_{sample}/(^{15}N/^{14}N)_{air}] - 1 \approx +100\text{-}300\ ^o/_{oo}$ (Wiens *et al.*, 1986); for the Martian atmosphere, $\delta_{air}^{15}N \approx +620\ ^o/_{oo}$ (Nier and McElroy, 1977; Owen *et al.*, 1977)), are difficult to produce by shock emplacement of gases from the surrounding rock.

This is so if only because the total nitrogen content in EETA 79001 glasses is, on the whole, not higher than that in the nonvitrified meteorite material. Indeed, according to Becker and Pepin (1984) and Wiens *et al.* (1986), in stepwise heating in an oxygen environment at $T > 700\ °C$, C1 glass releases ~0.45 ppm, C2 − ≈0.26 ppm, and C3 − ≈0.19 ppm nitrogen (at $T \geq 550\ °C$, respectively, 3.2 ppm, 1.1 ppm, and 3.5 ppm), whereas for the content in the surrounding rocks for EETA 79001 one obtains 0.13 (0.86) ppm, 0.48 (1.29) ppm (Becker and Pepin, 1984), 1.31 (28.0) ppm (Becker and Pepin, 1986), and for Shergotty, 0.13 (4.1), 0.79 (27.0) ppm (Becker and Pepin, 1966), and 2.3 (11.7) ppm (Wright *et al.*, 1986).

Thus there is rather an indication of the process going in the opposite direction, i.e. of the gases escaping from a rock heated and melted by impact compression and friction in the course of glass formation.

Heating and/or melting of the material and rearrangement of its crystalline structure should be accompanied by two processes: (1) degassing of the material and enrichment of the residual gas atoms and molecules in the molecular structure of the glass in heavy isotopes, and (2) redistribution of the residual gas atoms and molecules in the molecular structures of the glass solvent and in microvesicles, which likewise results in isotope fractionation.

Consider a few examples.

The enrichment of a rock by heavy isotopes (deuterium) in the course of nonequi1ibrium impact-induced degassing was shown to occur in serpentine by Tyburczy *et al.* (1990). Slow vaporization of 88% of molten $Mg_2SiO_2$ produces shifts in the residue of $\delta^{17}O = 21\ ^o/_{oo}$ and $\delta^{18}O \approx 43\ ^o/_{oo}$ (Davis *et al.*, 1990). Volatilization of ~70% $N_2$ from liquid magma may result in an increase of $\delta_{air}^{15}N$ in the residue of up to $+20\ ^o/_{oo}$ (Exley *et al.*, 1987). Anticorrelation between $\delta_{air}^{15}N$ and the total nitrogen content in magmatic rocks was pointed out more than once (Mayne, 1957; Sakai *et al.*, 1990), although is not always confirmed (Exley *et al.*, 1987). There may be many reasons for discrepancies, including inhomogeneity, strong degassing and, conversely, saturation by atmospheric gases of the upper mantle rocks, which (MORB etc.) are what is primarily accessible for investigation. Usually the nitrogen content in terrestrial magmatic rocks lies in the range ~1÷50 ppm (Mayne, 1957; Wlotzka, 1961; Becker and Clayton, 1977; Hoefs, 1987), while being possibly somewhat lower, ~1-2 ppm, in basalt glasses (Exley, 1987), whereas for tektites and other types of impact glass the figure 10-20 ppm is more typical (Murty *et al.*, 1989). In rocks, nitrogen is contained in the form of a physical solution or chemically bound ammonium $NH_4^+$ which replaces K, Na etc. metals, and of its derivatives, to wit, amine $NH_2$ and imine NH, replacing the hydroxyl OH (Mulfinger, 1966). The physical solubility of $N_2$ in glasses is ~0.2 ppm/atm at $T \approx 1,400\ °C$, whereas the concentration of chemically bound nitrogen in them under reducing conditions and at the same temperature may be ~$3\cdot10^4$ times higher. Oxidizing environment makes chemical binding practically impossible (by replacing NH by the hydroxyl OH) and reduces noticeably the physical solubility (down to ~0.13 ppm) (Mulfinger, 1966). The presence of water acts in the same direction. Apparently because of the presence of oxygen the fraction of atmospheric nitrogen trapped by natural glasses does not usually exceed a few percent of its initial value in them (Murty *et al.*, 1989), the abundance of nitrogen in basic rocks with enhanced water content (upper magma etc.) and in acidic rocks being relatively low (Murty *et al.*, 1989; Javoy and Pineau, 1986). Note that traces of water affect strongly even the character and completeness of Ar isotope extraction from rock samples in their stepwise heating (Zeitler and Fitz Gerald, 1986).

As known from the physics of shock waves (e.g., Baum *et al.,* 1975), a porous medium is heated by a passing shock wave stronger than a continuous one, which is supported also by the above experiments of Wiens (1988a,b) and Tyburczy *et al.* (1990). One may suggest that the regions which in EETA 79001 were impact-melted into glass had previously a higher porosity and, hence, a higher content of gases which could penetrate here through these channels also from other reservoirs (Wiens, 1988b) (we may recall in this connection the excess $^{129}$Xe concentration in the glasses). The initial concentrations of nitrogen in the rocks eventually turned to glass could be as high as ~10 ppm or more. Naturally, the somewhat enhanced nitrogen content prevailed before the impact also in the surrounding material which likewise underwent partial degassing when impact-heated to temperatures below the melting point.

When part of the nitrogen escapes under nonequi1ibrium conditions because of the fast impact-induced heating and decomposition of NH-groups, the rest becomes enriched in $^{15}$N, this enrichment in the case of ammonia occurring more efficiently (Sakai *et al.,* 1990) than it does in the diffusion of $N_2$ molecules (see earlier). As a result, the rest of the chemically bound nitrogen can be expected to contain an excess $\delta_{max}^{15}N \sim$ 10-100$^o/_{oo}$ or more.

Finally, it is hard to understand why the extraction of nitrogen in the course of stepwise heating should not in itself had to an ever increasing enrichment of the rest in the heavy isotope. Indeed, what we have said earlier (with the exception of the rate of the process) on the preferential escape at first of the light nitrogen should have a direct bearing on the stepwise heating of samples in the isotopic shift analysis. In particular, all the above authors, with the exception of Fallick *et al.* (1983), carried out nitrogen extraction in the presence of oxygen. It is this that apparently accounts for the fact that Fallick *et al.* succeeded in extracting by pure pyrolysis performed up to 1,200 °C on a Shergotty sample only 4.63 ppm N (with $\delta_{air}^{15}N$ = -14.8 $^o/_{oo}$), whereas Wright *et al.* (1986) recovered in combustion 11.6 ppm of the heavier nitrogen (with $\delta_{air}^{15}N$ = +28.6 $^o/_{oo}$). It can be suggested that even combustion in oxygen does not extract nitrogen completely, since neutron activation measurements of the nitrogen content in rocks yield frequently figures an order of magnitude higher than those obtained in stepwise heating (Becker and Pepin, 1986) (possibly because of a part of the ammonium left in the sample).

Further, while initially, at 500-550 °C < $T$ < 700 °C, normal (or even depleted in $^{15}$N) gas is released in large amounts, and we have seen (Wiens and Pepin, 1988; Murty *et al.,* 1989) that one can hardly consider this component an atmospheric contamination, later on its amount decreases strongly with increasing temperature, and the composition becomes enriched in $^{15}$N. This is valid, as a rule, for terrestrial samples (Exley *et al.,* 1987), Shergotty (Becker and Pepin, 1986; Wright *et al.,* 1986) and EETA 79001 (Becker and Pepin, 1986), and, to a still greater extent, for impact glasses of terrestrial rocks (Wiens and Pepin, 1988) and EETA 79001 (Becker and Pepin, 1984).

The glasses differ from crystalline rocks apparently in that the somewhat fractionated nitrogen still remaining in them divides subsequently into two components: (1) a less strongly bound and lighter component sited primarily in microvesicles, and (2) a component enriched in $^{15}$N which resides in intermolecular space or, more probably, is chemically bound in NH groups. It is in this way that one can interpret the crushing experiments of Wiens (1988a,b) on EETA 79001 glass and impact basalt glasses. If, however, one takes all fractions released for $T \geq$ 500 °C (or 550 °C), then the net nitrogen enrichment in the EETA 79001 glasses Cl, C2, C3 will be fairly modest, namely, $\delta_{air}{}^{15}N \approx$ 15-46-40 °/$_{oo}$, respectively. Such a shifts can be obtained if 70-90% N escapes in the course of glass formation. Then the initial nitrogen concentration in the glass-forming rocks should be only ~10-15 ppm, which are moderate and plausible values for terrestrial rocks.

Thus if we take into account (1) the "terrestrial" nitrogen released at 500-550 °C (and rejected), (2) possible isotope fractionation of nitrogen extracted in the course of analysis from the glass, which are processes deserving additional scrutiny, as well as (3) the practical impossibility of trapping of external gases by the glass formed in the impact, any relation of the SNC nitrogen with the anomalous nitrogen in the Martian atmosphere disappears.

## VI. ON A POSSIBLE NATURE OF THE OXYGEN ISOTOPE SHIFTS IN METEORITES

*6.1. State of the Art of the Oxygen Isotope Shift Problem for Meteorites*

Interpretation of the oxygen isotope shifts in SNC meteorites might seem at first glance to present a serious challenge. As a matter of fact, by Clayton and Mayeda (1983) (see also Clayton *et al.,* 1976; Begemann, 1980) when placed on the oxygen three-isotope diagram, $\delta^{17}O \div \delta^{18}O$, the SNC data lie parallel but slightly above the straight line with a slope $m = \delta^{17}O/\delta^{18}O = 0.52$ describing the mass-dependent equilibrium fractionation (of the type of diffusion etc.) of $^{17}O$ and $^{18}O$ in terrestrial or lunar conditions, whereas other basaltic achondrites (hypersthene achondrites and eucrites, including howardites, diogenites, mesosiderites and pallasites) are located a little below this line. Aubrites fit exactly onto the terrestrial straight line. Whence it follows (Clayton and Mayeda, 1983) meteorites are depleted in $^{16}O$ relative to the terrestrial abundance in SMOW (standard mean-oceanic water) by ~0.62°/$_{oo}$ (±0.1°/$_{oo}$), whereas eucrites, conversely, are enriched in $^{16}O$ by ~0.41°/$_{oo}$ (±0.1°/$_{oo}$). The isotopic ratios for oxygen in other types of meteorites are likewise described by a mass-dependent fractionation although the lines of their fractionation may sometimes be offset noticeably parallel to the terrestrial line. It might appear that the low-abundant heavy isotopes ($^{17}O$, $^{18}O$) obey the laws of mass fractionation, whereas the main isotope, $^{16}O$, does not.

The behavior of $\delta^{17}O \div \delta^{18}O$ for the high-temperature phases of carbonaceous chondrites of the C2-C3 type different. It represents a straight line with a slope $m \approx 1.0$ lying within $\delta^{17}O \approx \delta^{18}Q \approx -40°/_{oo} \div +10°/_{oo}$. Clayton *et al.* (1976, 1973) attributed all deviations from the terrestrial line to initial inhomogeneities in the protosolar nebula (see also Begemann (1980)), in particular, to the formation of high temperature phases of C2-C3 in a region originally enriched in O in the course of nucleosynthesis which at the same time did not involve such abundant light elements as Mg, Al, Ca, Ti. For conventional chondrites the parallel displacement (by at least ~2°/$_{oo}$) of their line relative to that of terrestrial fractionation was ascribed to a reduced relative abundance of the $^{16}O$ component, whereas for the hydrated matrix of C2 carbonaceous chondrites the relative abundance of $^{16}O$ should, on the contrary, be increased by ~4°/$_{oo}$.

The idea of the presence of two (or even three (Rubin, et *al.,* 1990) or four (Begeman, 1980)) phases, namely, a "normal" and an $^{16}O$-enriched phase, was further developed in the concept of reservoirs and/or parent bodies for SNC meteorites (supposedly of Mars) and, e.g., eucrites, or for the Earth and Moon, or even for different components of the same meteorite (Rubin *et al.,* 1990).

However the arguments of Clayton *et al.* (1973) for two independent oxygen reservoirs are not convincing for at least two reasons.

First, the expression derived by the above authors for evaluating the fraction of excess $^{16}O_{excess}$ (Clayton and Mayeda, 1983), contains an error. This expression is as follows

$$^{16}O_{excess} = \frac{0.52}{1-0.52}\delta^{18}O - \frac{0.52}{1-0.52}\delta^{17}O \tag{4}$$

where $0.52 = \delta_0^{17}O/\delta_0^{18}O$ corresponds to the line of equilibrium fractionation of oxygen of "normal" (terrestrial, lunar etc.) composition, and $\delta^{17}O$ and $\delta^{18}O$ are shifts in oxygen containing an admixture of pure $^{16}O$, so that (4) can be rewritten

$$^{16}O_{excess} = \frac{\delta_0^{17}O \cdot \delta^{18}O - \delta_0^{18}O \cdot \delta^{17}O}{\delta_0^{18}O - \delta_0^{17}O} \tag{5}$$

Relations (4) and (5) do indeed describe a straight line parallel to the equilibrium fractionation line with a slope of 0.52 on the three-isotope diagram. However the correct expression for $^{16}O_{excess}$ is

$$^{16}O_{excess} = \frac{\delta_0^{17}O \cdot \delta^{18}O - \delta_0^{18}O \cdot \delta^{17}O}{\delta^{17}O - \delta^{18}O} \tag{6}$$

or

$$^{16}O'_{excess} = \frac{0.52 - m}{1 - m} \cdot \delta_0^{18}O \tag{7}$$

or again

$$^{16}O''_{excess} = \frac{1 - \dfrac{m}{0.52}}{1 - m} \cdot \delta_0^{17}O \tag{8}$$

where $m = \delta^{17}O/\delta^{18}O$ is the ratio of the measured oxygen shifts in the given rock.

We immediately see from (7) and (8) that (a) different points on a straight line parallel to the equilibrium terrestrial fractionation line have different values of $^{16}O_{excess}$, and (b) for $m \neq 0.52$ no definite value of $^{16}O_{excess}$ can be attributed to a rock at all, since it is unclear whether one should start from $\delta_0^{18}O = \delta^{18}O$ or $\delta_0^{17}O = \delta^{17}O$, so that the concept of oxygen mixing from two reservoirs of normal and $^{16}O$-enriched composition loses to a considerable extent original meaning.

Second, it is known (Matsuhisa *et al.*, 1978) that it is often difficult in real conditions to reach complete isotopic equilibrium in a finite time, particularly at low temperatures. Therefore the observed isotopic shifts are not infrequently due to kinetic isotopic effects.

*6.2. Kinetic Effects of Isotope Fractionation in Gases and Solids*

More natural appears, however, another approach which relates the various isotopic features observed in various objects to the process proper of their formation and evolution rather than to a combination of some *a priori* mysterious initial conditions (Nosik, 1986).

In the recent years, investigation of mass-independent nonequilibrium processes of isotope fractionation capable of producing the various observed shifts in an initially standard isotopic mixture has been acquiring particular attention.

Such are the photochemical methods of gas phase separation of the isotopes (e.g. Letochov, 1978. 1983) of carbon, nitrogen, chlorine, mercury whose study was started as far back as the '30s. They are based on the use of resonant radiation, its frequency depending noticeably on nuclear mass. Considerable interest was attracted to the work of Basov *et al.* (1974) and, particularly, of Thiemens and Heidenreich (1983) on the enrichment by heavy isotopes of nitrogen oxides and ozone, respectively, frozen out of the gases subjected to electric discharge. The latter authors directly related the results of their experiments with the above mentioned oxygen anomalies in the high temperature phases of carbonaceous chondrites, although the actual fractionation mechanism is not completely clear (Heidenreich and Thiemens, 1986; Bates, 1988).

Studies of mass-independent isotope fractionation in the solid condensed phase have only recently started, although the few available publications on this subject suggest the possibility of manifestation of strong effects. Indeed, in the partially deuterated synthetic MgO the probability of formation of $D_2$ is about four times that of $H_2$ (Freund *et al.,* 1982). A nondiffusional mechanism of an isotope extraction in stepped rock heating which is capable of distorting the dating by the $^{40}Ar/^{36}Ar$ method was pointed out by a number of authors (e.g.., Zeitler and Fitz Gerald, 1986; Dallmeyer, 1982). Interestingly, the presence of water in trace amounts plays an essential role in these processes (we should recall here the differences in $^{15}N$ extraction in pyrolysis and combustion discussed in Sec.5.2.3); for an analysis of the effect of water on the mobility of hydrogen and oxygen in quartz and feldspar see Elphick and Graham (1988).

Nosik (1986) was one of the first to systematize the influence of kinetic effects on the isotopic shifts of sulfur, carbon and oxygen in the rock-formation processes. He pointed out that the kinetic isotopic effect is always larger than the thermodynamic one, which accounts for the fact that newly formed compounds are enriched in light isotopes, and close to equilibrium they are always characterized by $m = \delta^{17}O/\delta^{18}O > 0.5$, whereas in the remainder $m < 0.5$; the stronger oxidized compounds contain a larger amount of the heavy isotopes, so that the ratio $m$ in the mineral decreases with increasing oxidizing capacity of the mineral-forming medium, and *vice versa*. Basing on the measurements of $\delta^{17}O$ and $\delta^{18}O$ in over 250 terrestrial minerals and rocks, Nosik came to the conclusion that $m$ may deviate from the equilibrium value of $\approx 0.5$ to any side from $-\infty$ to $+\infty$. Nevertheless, for most of the minerals $m$ lies between 0.5 and 1.4, which corresponds (according to his criterion) to 100% and 71% of the equilibrium level, respectively. On the average, the formation of terrestrial minerals occurred at a level of 55% to 92% to equilibrium, that of high-temperature anhydrous phases of C2-C3 meteorites studied by Clayton *et al.* (1973, 1976) at ~75% ($m = 1$), and of lunar samples - at 67% to 75%. An informative illustration is provided by the distribution of $m$ in solidified acidic lavas; namely, in the depth range 0-50 m, the value of $m$ varies within $-2 \leq m \leq +2.4$ (and even $+8.85$), approaching most closely ($m \approx 0.3-0.9$) the equilibrium value (0.52) in middle regions (Boyarskaya *et al.,* 1990). Such a distribution is attributed by the authors to uncompleted exchange with the water penetrating into rocks after their eruption.

Considering that the composition of SNC meteorites is close to that solidified lavas (however, significantly, of ultrabasic ones), one may also expect in their bulk, within a few tens or hundreds of meters, deviations of $m$ from 0.52, primarily toward larger values (basic rocks!), so that it is of these meteorites that are typical the values of $m$ which are reached in most cases. In this context one cannot, for instance, maintain that the Brachina does not belong to the SNC group basing simply on the differences in the $\delta^{17}O/\delta^{18}O$ ratio, as this was done by Clayton and Mayeda (1983), since different layers in a rock may have, as we have seen, $m \gtreqless 0.5$.

We see that just as before (Sec. 5.1), when the enhanced content of iron in SNC meteorites was considered, the data on the oxygen isotopes lead one to a conclusion that their parent body was ejected from the Earth from an exotic geochemical province of ultrabasic volcanism, so that the characteristic size of this body could have been only ~100 ($\pm 1$ dex) m. From this standpoint it would be extremely important to study in detail the oxygen isotope composition of the corresponding terrestrial rocks (which would possibly rehabilitate also Brachina).

**VII. HYPOTHESIS OF AN ELECTROLYTIC NATURE OF THE ISOTOPIC SHIFTS**

*7.1. On the Possibility of Isotope Fractionation in Solid Electrolytes*

Turning back to nonequilibrium (kinetic) effects in the separation of oxygen isotopes, we should like to note that in a number of minerals (for instance, oxides of the type of $CeO_2$, $ZrO_2$, $ThO_2$, $Bi_2O_3$) with the regular of somewhat disturbed cubic fluorite structure, $CaTiO_3$-type perovskites (Subbarao and Maiti, 1984), mullite-type aluminosilicates $(1.5-2)Al_2O_3 \cdot SiO_2$ (Meng and Huggins, 1984) the oxygen anion $O^{2-}$ exhibits a fairly high mobility, thus conferring to such minerals the properties of electrolytes. Since the passage of electric current is an essentially nonequilibrium process, one may expect it to be capable of producing isotopic shifts as well.

Such solid electrolysis find an application in electric-fuel battery cells, as oxygen concentration sensors etc. Suitable for technical applications are materials with electric conductivities $\sim 10^{-2} \div 1$ mho/cm at temperatures of hundreds of deg °C, the conduction being provided primarily by the mobility of the ions involved, in this case of $O^{2-}$.

The oxygen-based ($O^{2-}$) electric conductivity of solid minerals is due to the presence of vacancies in a continuous network made up of oxygen ions. The vacancies originate from deviations from stoichiometry, introduction of dopants, etc.

In real life, many more materials possess a weaker or mixed ionic-electronic conduction, which makes their use in applications hardly likely. Different ionic conduction mechanism may frequently operate simultaneously or replace one another; for instance, the protonic and oxygen (or, say, $Na^+$-based) conductions may compete, the actual type of conduction depending on the temperature, oxygen or water vapor pressure, doping by some other cations, method of preparation used and other, frequently poorly controlled parameters. Interestingly, there is a relation, still not well understood, between the oxygen-based electric conduction of complex oxides and the possibility of their becoming high-$T_c$ superconductors (Sholhorn, 1989).

It thus becomes clear that practically all rock-forming minerals with the number of oxygen atoms exceeding by at least a factor 1.3-1.4 that of all other atoms should exhibit to some extent the oxygen-based component of electric conductivity.

On the other hand, electric transport involving the motion of anions should reveal noticeable isotopic effects. The experiments of Hainovsky *et al.* (1986) showed that protonic conduction in $CsDSO_4$ may excess by two orders of magnitude the deuteronic one. Sherban *et al.* (1988, 1989) found that the conductivity of $SrCeO_3$-type perovskite (+5%$Yb^{3+}$) charged by hydrogen is 2.5 times higher than that of the same material charged by deuterium, "which is significantly greater than the classical value of $2^{1/2}$" if the electric conductivity is due to the mobility of $H^+$ or $D^+$. The first case originates apparently from differences in the tunneling effects, while in the second case we have a difference $\Delta E = 0.04\text{-}0.055$ eV between the activation energies of their ground state (it should be compared with the total binding energy, ~0.5-2 eV (Norby, 1990)).

One can easily imagine that other, less well studied effects, for instance, of the cooperative type, are capable of producing differences in the ion mobilities of different isotopes. Indeed, there are indications that in some cases alkali ions move in a correlated (pairwise) way. It brings about the so-called "polyalkaline effect", where combined ions of different mass (e.g. K and Na) reduce the electric conductivity of a solid electrolyte of mixed composition below that of the electrolyte containing ions of one type only (Ingram *et al.,* 1980). A "caterpillar" effect has also been observed (Kobayashi *et al.,* 1990) where an ion can be forced out of a vacancy by another one hopping over to it.

Correlated transfer of many particles simultaneously is also possible (Dietrerich, 1990). The overall net-like pattern of possible ion motion in the direction of the imposed field, the presence or absence of obstacles, barriers etc. on the way may become essential (Ingram *et al.,* 1990). Apart from the lightest $H^+$ and $D^+$ ions, the non-classical behavior of the mobility ratio was pointed out for $^6Li^+$ and $^7Li^+$ ions also (Jain and Peterson, 1982).

In this context the oxygen case acquires significance, where the heavy isotopes $^{17}O$ and $^{18}O$ act as small impurity compared with the abundant $^{16}O$ atoms. Indeed, it is known that the properties of a monoisotopic crystal lattice may differ strongly from those of a lattice containing other isotopes as well. A remarkable illustration is provided by the pure $^{12}C$-diamond whose heat conductivity is ~1.5 times higher than that of the natural material containing only 1.1% of $^{13}C$. Because of their high abundance, the $^{16}O^{2-}$ anions form in a solid electrolyte a continuous network (~ one-dimensional lattice) with rare inclusions of single anions $^{17}O^{2-}$ (0.0375%) and $^{18}O^{2-}$ (0.1995%). The overall situations for $^{16}O$, on the one hand, and $^{17}O$ and $^{18}O$, on the other, are radically different from the topological viewpoint. Therefore it can be assumed that the exchange mobility of the $^{16}O^{2-}$ ions along a chain of identical particles should excess noticeably the diffusive mobility of the $^{17}O^{2-}$ (or $^{18}O^{2-}$) relative to $^{16}O$. (The same can apply to the mobility of oxygen in ice, although the electric conductivity in it is primarily of protonic nature.)

This possible "topological" mechanism of isotope separation in solid electrolytes has not, to our knowledge, been suggested or investigated up to now. Neither has been studied the possibility of isotopic effects similar to the "polyalkaline" behavior. At the same time revealing them should not apparently cost much effort if one uses, for instance, the well known electrochemical oxygen pump which requires application of an external voltage across a wall of solid electrolyte (Subbarao and Maiti, 1984). It can be expected that the isotopic shifts in a gas pumped through such a wall with oxide-based electric conductivity will not be described by mass-dependent, purely diffusive relations.

Electrolytic transport may play an essential role in geochemical processes, for example, in the formation of heavy oxidized iron ores (Morris *et al.,* 1980). Therefore investigation of the presence and of specific properties of oxygen isotopic shifts in such ore bodies could probably shed light on the nature of such anomalies in same meteorites.

*7.2. Hypothesis of an Electrolytic Nature of Oxygen Isotopic Shifts in Meteorites*

If the mobility ratios in rocks (and in ice) between $^{16}O^{2-}$, on the one hand, and $^{17}O^{2-}$ and $^{18}O^{2-}$, on the other, differ noticeably from the classical values, it appears only natural to try to attribute the shifts in the three oxygen isotope diagram revealed by Clayton *et el.* (1973, 1976) to some specific nonequilibrium process which accompanied the formation of the various meteorite types. Since meteorites are fragments of asteroids, and most of the latter, as we believe, represent fragments of a planet with $M \leq 0.5 M_{moon}$ broken up because of the explosion of the electrolysis product-saturated ices of its envelope (Drobyshevski, 1986), electrolytic transport could have played a major role in the formation of isotopic shifts in them.

We are far from the idea of being able to present here the final explanation of the nature of the oxygen isotope anomalies in the various types of meteorites; indeed, to do this, one would have to offer a specific scenario of the formation of the rocks they are made of as this was done for a number of terrestrial cases by Nosik (1986). By the way, one could not totally exclude in the analysis the effect of inhomogeneities in the protosolar nebula postulated by Clayton *et al.* (1973), or of nonequi1ibrium initial conditions which were assumed to have existed in it by Thiemens and Heidenreich (1983).

It should be noted that the motion of $O^{2-}$ ions through a crystal lattice made up primarily of $^{16}O$ would result in a buildup in it of the slower moving $^{17}O^{2-}$ and $^{18}O^{2-}$ ions, the lattice acting as a sieve letting through easily the $^{16}O^{2-}$ ions. The accumulation (or depletion) occurs, however, only up to a certain (stationary) limit associated with the actual boundary conditions (type of mineral, including, say, metals or ice), together with the electrochemical processes in their vicinity, such as the electrolysis of ice, oxidation or reduction of chemical compounds (the appearance of L, LL and H components in meteorites) etc. Therefore some sections of the electric circuit in the current path can be depleted in $^{16}O$ compared with $^{17}O$ and $^{18}O$ (say, L, LL and H chondrites) and others, enriched in it. Among the latter can possibly be ice, as well as small dispersed grains of the C2 matrix trough which current does not pass, since due to their small size the potential difference across each of them is not high enough for the charge carriers to penetrate through the double electric layers created at their interfaces with ice. Therefore the isotopic composition of oxygen in them is determined primarily by diffusion from the surrounding ice. It is fragments of such ice with organic impurities and inclusions of mineral grains which contain also the products of its electrolysis that are the cometary nuclei (Drobyshevski, 1988). They are ejected into space by exploding envelopes of bodies like Ganymede (Drobyshevski, 1989) or Titan (Drobyshevski, 1981).

Thus one can imagine that as current flows through an icy moonlike body, bulk electrolytic decomposition of ice and accumulation in it of the electrolysis products should be accompanied by the formation from the various components of this body of parent bodies of the future meteorites, which would be depleted or enriched in $^{16}O$.

*7.3. A Scenario of the Formation of Oxygen Isotopic Shifts in the Ejection of SNC Meteorites*

We are now in a position to suggest a mechanism for the formation of the observed isotopic shifts different from the possibility of nonequi1ibrium terrestrial enrichment of the rocks of SNC meteorites in the $^{17}O$ and $^{18}O$ isotopes considered at the end of Sec.6.2.

The rock fragments ejected into space in a hypervelocity impact are accelerated by a jet of very hot gas whose major component is oxygen composing initially the vaporized target rock material. The gas pressure on one side of the fragment is $\sim 10^2$ kbar, while on the other it may be orders of magnitude lower.

The SNC fragment rock can be somewhat conventionally represented as a mixture of the heavy cations of Si, Mg, Fe, Ca, Al (arranged approximately in the order of their decreasing content) incorporated into a more or less continuous (because of their comparatively large content) network of the oxygen anions $O^{2-}$ of $\sim$one half the former in weight. The main force acting against the accelerating gas pressure on a fragment is that of inertia. Because of their lower mass, the oxygen ions would tend to break away from the heavier metal cations, to outdistance them, were it not for the chemical bonding. Since, however, the oxygen ions are built into a continuous network, they possess, as we have seen (Secs.7.1, 7.2) a certain mobility with respect to the cations, thus mobility for $^{16}O^{2-}$ being possibly noticeably higher compared with that of $^{17}O^{2-}$ or $^{18}O^{2-}$ than this follows from their mass ratio. As a result, the oxygen atoms from the accelerating gas compressed to $\sim 10^2$ kbar filtered through the fragment under acceleration, passing preliminarily through a layer of molten rock which forms between the surface of the fragment and the hot gas flowing around it. Since the rate of filtration of $^{17}O^{2-}$ and $^{18}O^{2-}$ through the fragment is noticeably less than that of $^{16}O^{2-}$, they will build up in the rocks of the fragment, thus producing the observed relative deficiency of $^{16}O$.

It should be pointed out that the patterns of $O^{2-}$ passage through the crystal lattice when being transported by electric current under an applied electric potential difference and in the case of filtration under a high pressure may be different. In the first case, the $O^{2-}$ ions will move individually in sequential hops determined primarily by the presence of oxygen vacancies and other defects. In the second, one can imagine the abovementioned (Sec.7.1) cooperative mechanism of hydrodynamic type capable of displacing whole groups of ions bound in a chain and made up of $^{16}O$.

Thus this approach to the interpretation of the oxygen isotope ratio shifts in SNC meteorites assumes them to appear in the process and as a result of their gasdynamic ejection from a parent planet.

**VIII. GENERAL CONCLUSIONS**

Thus a more careful inclusion of the parameters determining the ejection from planetary bodies into space of rock fragments in hypervelocity impact events has brought us to a fairly unexpected conclusion, to wit, that meteoroid impacts of energy $W \geq 10^6$ Mt TNT equivalent may result in ejection from the Earth of bodies of up to $\varnothing \sim 1$ km, i.e. asteroids! Note that we have taken into consideration (1) the increase of rock strength up to $s_{max} \approx 100$ kbar under high confining hydrostatic pressure, and (2) the jet nature of effective acceleration path length $S_{eff}$ nearly threefold, so that $S_{eff} \approx D/3$.

It should be mentioned that greatly overabundant number of smaller fragments ($\varnothing \leq 10^2$ m) (Rabinovitz, 1991) can escape from the Earth in impacts of $W \geq 10^3$ Mt TNT as well, however such events should not result in global ecological catastrophes and a change of geological epochs on our planet. The processes occurring in continental impacts with $W \sim 10^6$ Mt TNT represent apparently a bordering case in its climatic and ecological consequences for two reasons: (1) the impact itself creates a full-scale global nuclear winter thus causing a mass extinction, and (2) it results in an ejection into space of numerous sufficiently large fragments whose partial infall back onto the Earth during the subsequent 1–3 Myr would produce a series of "minimal" nuclear winters. Interestingly, such impacts with $W \geq 10^6$ Mt TNT occur, on the average, once every 25-30 Myr, which coincides with the intervals between large-scale extinctions of the biota, the falls of returning large fragments providing the shower-like bombardment necessary to account for the stepwise character of some extinctions. The ejection of km-size fragments in hypervelocity impacts solves the problem of the origin of secondary craters with diameters of up to $D \approx 30 \div 40$ km on the Moon and similar bodies.

Together with a possibility of the shower bombardment of the Earth by fragments from fairly infrequent explosions of electrolyzed remote ice bodies (Drobyshevski, 1989), the above mechanism is capable of removing the difficulties associated with the large spread of trace element content in the boundary layers between different geological epochs, if one takes into account the consequences of catastrophic impacts on the Moon. One could possibly make an assumption, however extreme, that the transitions between the geological epochs are always caused by large-scale meteoroid impacts. Besides shedding light on the causes of global variations of the climate and biota, the possibility of the impact-induced ejection of very large boulders from the Earth and, naturally, from other terrestrial-type planets into space permits one to explain the nature of many NEAs, which are traditionally known to be deficient in their possible sources. It opens up also a new approach to the solution of the intriguing problem of the origin of eight basalt "martian" meteorites, the shergottites, nakhlites, and Chassigny (SNC). It turns out that they can be of terrestrial rather than Martian origin, impacts with $W \geq 10^3$ Mt TNT being strong enough to eject them into space. Such an assumption removes immediately a number of contradictions and difficulties, namely (1) the lunar-to-Martian meteorite mass ratio opposite to the expected figure (presently it is 0.025 whereas it should be 2,500); (2) the absence on Mars of sufficiently young ($\leq$ 200 Myr) large-diameter crater; (3) the presence in the EETA 79001 shergottite of a remarkable high content of organic matter. The latter may be of significance for the panspermia ideas.

At the same time there are points which still remain unclear and require resolution. However the new approach opens up new possibilities here too. The unclear aspects relate primarily to the chemical composition of SNC meteorites and isotopic ratio shifts of noble gases, nitrogen and oxygen in their material.

In their composition, all SNC meteorites correlate with terrestrial rocks of the so-called basaltic flood volcanism characterized by a high iron content. Therefore as long as no other terrestrial samples are found among meteorites, one has to assume that all the meteorites combined in the SNC group originate from the same ejection event. Also, since, judging from the shock-induced metamorphism, the ejection occurred ~200 Myr BP, and other events of this kind should have since occurred, one has to accept that these meteorites are very likely also fragments of the same body of at least a few tens of meters in size which disintegrated several million years ago.

It thus becomes clear that a search among Antarctic meteorites for terrestrial samples of another type is an urgent problem; also, among the finds there should be acid and sedimentary rocks, some of them with remains of terrestrial fossil organisms.

On the whole, the Ar/Kr/Xe ratios in SNC meteorites, just as the proportions between their main isotopes, are hardly distinguishable from those typical of terrestrial rocks. Nevertheless, it is believed that the isotopic ratio shifts observed for the noble gases and nitrogen in SNC glasses reflect the anomalies revealed in the Martian atmosphere (this relates to the enhanced ratios

$^{15}$N/$^{14}$N, $^{40}$Ar/$^{36}$Ar, $^{129}$Xe/$^{132}$Xe), although shock wave experiments have demonstrated the low efficiency of atmosphere gas trapping by the rocks.

On the other hand, the $^{40}$Ar/$^{36}$Ar ratio in terrestrial basalt varies from 296 (air) to ~$10^5$, a range that includes the values observed in SNC rocks (up to ~2,000). The inclusion of a possible isotope fractionation of nitrogen in the course of its release by various mechanism from terrestrial rocks containing typically ~10÷15 ppm N$_2$ is capable of accounting for the observed mean excess in $\delta_{air}^{15}$N ~ +10$^o/_{oo}$ ÷ +50$^o/_{oo}$ (and even more in some high temperature fractions under stepwise heating).

As for the $^{129}$Xe/$^{132}$Xe ratio, in terrestrial rocks it is known to vary from 0.983 (air) to 1.15, in the Martian atmosphere it is 2.5(+2,-1), and in some high temperature Xe fractions from EETA 79001 glasses it reaches ~2.4. The coincidence of the latter value with the Martian figure (while not very reliable) is certainly impressive and thus one cannot reject a possibility for at least one of SNC meteorites to be of Martian origin, indeed. The excess of $^{129}$Xe is assumed to be due to the radiogenic isotope produced in the decay of $^{129}$I. The I/Br ratio in EETA 79001 is indeed high, but it is usually attributed to terrestrial contamination. Bearing in mind the strong heterogeneity of the Earth's mantle and the presence in it of numerous and, quite frequently, small-scale reservoirs with the relic original (or fissiogenic) composition of elements, one can readily suggest that it is the composition of one of such anomalous reservoirs that the EETA 79001 glasses represent. Measurements of $^{129}$Xe content in terrestrial rocks reveal a continuous increase of the maximum value of the $^{129}$Xe/$^{132}$Xe ratio, suggesting the possibility of finding in the future rocks with $^{129}$Xe/$^{132}$Xe > 2. Apart from this, one has to keep in mind that the processes of impact-induced rock degassing and of subsequent isotope fractionation which only recently have begun attracting attention may provide unexpected explanation for the origin of a number of isotope anomalies.

The 15 year old history of revealing in various groups of meteorites various types of oxygen isotopic shifts, when these groups plotted on a three-isotope diagram either lie parallel to the terrestrial fractionation line or fit onto lines with a different slope, has not yet led to a full understanding of the nature of the observed anomalies. And although several mass-independent mechanisms of isotope fractionation have been pointed out, and numerous examples have been given of nonequi1ibrium fractionation of the isotopes of carbon, oxygen and sulfur in terrestrial rocks which are plausibly interpreted in terms of kinetic processes (Nosik, 1986; Boyarskaya *et al.,* 1990), nevertheless the most widely accepted opinion still remains the original suggestion of Clayton *et al.* (1973) of the existence of several reservoirs enriched or depleted in the $^{16}$O isotope, and it is these reservoirs from which the various groups of meteorites or their parent bodies formed. It is believed, for instance, that the parent body of SNC meteorites was depleted in $^{16}$O by ~0.6$^o/_{oo}$.

Starting from the opposite premise, namely, that the oxygen shifts originate from processes that accompanied the meteorite formation, we suggested, in addition to the simplest assumption of the oxygen isotope shifts observed in SNC meteorites being due to an initial nonequi1ibrium kinetic process in ultrabasic volcanic rocks on the Earth itself, we have put forward a hypothesis that these anomalies can be accounted for by different and, as shown by experiment, frequently nonclassical ion mobilities of different isotopes (e.g. of hydrogen and lithium) in solid ionic conductors. Most of the rocks where oxygen is the predominant species possess, besides the electronic, also oxygen-based electric conductivity. Note that the $^{16}$O$^{2-}$ ions form a continuous network whereas the $^{17}$O$^{2-}$ and $^{18}$O$^{2-}$ ions representing a small impurity act as discrete inclusions. Therefore when electric current flows through the icy envelopes of moonlike bodies resulting in the bulk electrolysis of ice (Drobyshevski, 1980, 1986), which leads to their eventual explosion and production of minor bodies (including meteorites), the $^{16}$O isotope undergoes redistribution relative to $^{17}$O and $^{18}$O in different components of the original body (ice, its mineral matrix, rocks, iron etc.).

Similar processes occur also in the gasdynamic acceleration of rock fragments by hot gases whose major component is oxygen liberated in the vaporization and thermal decomposition of rocks in the hypervelocity impact. The crystalline structure of a fragment acts as a sieve through which the $^{16}O^{2-}$ passes fairly easily along the continuous $^{16}O^{2-}$ chains driven by the $\sim 10^5$ atm pressure drop whereas the $^{17}O^{2-}$ and $^{18}O^{2-}$ ions get stuck and build up in it.

The proposed electrolytic mechanism requires naturally experimental verification. If it is confirmed, then the oxygen anomalies of different types of meteorites will indicate the major role played in their prehistory by electrolytic ionic processes in the solid phase. There are grounds to suggest that, in contrast to SNC meteorites, Martian rocks have, on the average, the normal oxygen isotopic composition, and that the line of its equilibrium fractionation should not differ from that typical of the Earth or Moon.

## ACKNOWLEDGEMENT

The author expresses sincere gratitude for valuable advice to Dr. W.M.Napier who, in support of the possibility of large fragment ejection, pointed out the significance of the large-size secondary craters, and possible consequences of the large boulders ejection possibility to the panspermia ideas. The author acknowledges also a stimulating discussion with Prof. L.K.Levskii of a number of aspects associated with the isotope ratio shifts.